# BLAST: A Wafer-scale Transfer Process for Heterogeneous Integration of Optics and Electronics

Yanxin Ji[1], Alejandro J. Cortese[2], Conrad L. Smart[3], Alyosha C. Molnar[1,2,4], and Paul L. McEuen[2,3,4,*]

[1]Department of Electrical and Computer Engineering, Cornell University, Ithaca, NY 14853, USA

[2]OWiC Technologies, Ithaca, NY 14853, USA

[3]Laboratory of Atomic and Solid-State Physics, Cornell University, Ithaca, NY 14853, USA.

[4]Kavli Institute at Cornell for Nanoscale Science, Cornell University, Ithaca, NY 14853, USA.

*Corresponding Author: Paul L. McEuen.    plm23@cornell.edu

**Abstract**

We present a general transfer method for the heterogeneous integration of different photonic and electronic materials systems and devices onto a single substrate. Called BLAST, for Bond, Lift, Align, and Slide Transfer, the process works at wafer scale and offers precision alignment, high yield, varying topographies, and suitability for subsequent lithographic processing. We demonstrate BLAST's capabilities by integrating both GaAs and GaN µLEDs with silicon photovoltaics to fabricate optical wireless integrated circuits that up-convert photons from the red to the blue. We also show that BLAST can be applied to a variety of other devices and substrates, including CMOS electronics, vertical cavity surface emitting lasers (VCSELs), and 2D materials. BLAST further enables the modularization of optoelectronic microsystems, where optical devices fabricated on one material substrate can be lithographically integrated with electronic devices on a different substrate in a scalable process.

**MAIN TEXT**

Heterogeneous integration of dissimilar materials—combining devices or circuits made from two or more functional materials in a single integrated platform, push technology advances in ways that single material systems cannot[*1-3*]. One of the most compelling examples is the integration of III-V photonic devices with silicon electronics, with applications ranging from on-chip high bandwidth communication[*4,5*] to wearable/implantable devices for diagnostics and therapeutics[*6-8*]. One approach to heterogeneous integration of dissimilar materials is heteroepitaxy, but it is a major challenge to grow different high-quality crystalline materials on a single substrate. While some notable successes have been shown[*9*], in general lattice mismatch and different thermal expansion coefficients prevent monolithic growth of defect-free material[*10*].

An alternative is transfer-based heterogeneous integration—the separate fabrication of devices on two or more substrates, followed by a transfer step to combine them. This allows growth and fabrication processes associated with each substrate to be independently optimized[*11*] and the two parts combined only at a later step. An ideal transfer method would provide: 1) high alignment accuracy, 2) high transfer yield, 3) wafer-scale transfer, 4) applicability to structures of varied dimensions and geometries, and 5) be amenable to subsequent lithographic processing. Within the past decade, a wide variety of transfer methods have been developed, such as elastomer stamp transfer[*12-14*], flip-chip bonding[*15,16*], pick-and-place[*17*] and fluidic assembly[*18*]. However, none of them fully meet the requirements above. For example, flip-chip bonding does not allow for subsequent processing, pick-and-place is a serial, slow process, and fluidic assembly does not work for small samples. Elastomer stamp transfer comes the closest and is currently being incorporated into commercial foundry process flows[*19*]. However, it requires suspending the devices to be transferred, is difficult for materials grown on chemically inert substrates such as GaN on sapphire [*20, 21*], and it typically operates at less than full wafer scale.

In this letter, we present BLAST, a robust, wafer-scale transfer technique for heterogeneous integration that meets all of the requirements listed above. Figure 1 illustrates the process, showing the heterogeneous integration of silicon electronic devices (Fig. 1A) and optical III-V µLEDs (Fig. 1B). We demonstrate transfer both GaAs-based and GaN-based µLEDs from their native substrates (GaAs and sapphire) onto a target Si substrate with precision alignment (~1µm) and high yield (~99.9%). We use this process combined with subsequent lithographic processing to fabricate broadband optical upconverters (Fig. 1C-F) that absorb low-energy photons (Si PVs) and emit higher-energy ones (GaN µLEDs). We finally show that this transfer technique can be used for a wide variety of materials/substrate combinations including the integration of VCSELs and foundry CMOS as well as for the layer-by-layer aligned assembly of 2D van der Waals materials.

BLAST consists of four steps: (i) *Bonding* of the µLED native substrate to a transparent sapphire carrier wafer, then (ii) *Lifting* of the µLEDs from the native substrate, then (iii) the *Alignment* of the µLEDs to features on the target substrate, and finally (iv) *Slide Transfer* of the µLEDs from the carrier wafer to the target substrate using a thermal slide technique. This is shown for the case of µLEDs in Fig. 2. The µLEDs are first fabricated on their native substrate using standard lithographic processes (see Materials and Methods), along with precision alignment marks. For GaN µLEDs, a metal layer is added for laser shielding during the removal process, as discussed in Materials and Methods. Separate, complementary alignments marks are also patterned on the target substrate for precision alignment. In most cases, a 500nm layer of SU8, a standard epoxy-based adhesion layer, is spun onto the target substrate.

For *Bonding* of the native substrate to the carrier wafer, two different polymer layers are applied, each playing a different role. Poly (methyl methacrylate) (PMMA) is spin-coated onto the native substrate as a protection layer to prevent damage to the µLEDs during etching and transfer. Separately, a layer of polypropylene carbonate (PPC), a thermoplastic polymer, is spin-coated onto the sapphire carrier wafer. PPC's role is to temporarily bond the carrier wafer to the µLED substrate. This is accomplished using a hot press at $p = 1.5$ bar and $T = 135°$ C. Both PMMA and the PPC are highly transparent (as is the sapphire carrier wafer) and not light sensitive, facilitating optical alignment. Both can be subsequently easily removed by standard processes.

The μLEDs are then *Lifted* from the native substrate, leaving behind only the μLEDs and alignment marks attached to the carrier wafer by the PPC (Fig. 2A). Removal of the devices from the native substrate can be accomplished by a variety of means. Here we use selective wet etching of the substrate in the case of GaAs μLEDs[22], stopping on an AlGaAs release layer. For GaN μLEDs, we use excimer laser lift-off[23] (Fig. 2A) that decomposes a thin layer of GaN to metallic gallium and nitrogen gas to release the devices.

For *Alignment*, the μLEDs on the carrier substrate are spatially (*x, y*) and rotationally (*θ*) aligned to pre-patterned alignment marks on the target Si substrate using a standard contact (ABM) mask aligner, as shown in Figs. 2A, 2B. They are brought into contact and heated to 100°C using a custom heater stage to promote adhesion. They are then removed from the aligner.

For *Slide Transfer*, the stack is further bonded using a hot press at $p = 0.7$ bar and $T = 155°C$, above the melting temperature of the PPC (~100-120°C), but below that of the PMMA(~190°C)[24,25]. Finally, they are separated by placing on a hot plate at $T = 160°C$ and using the lateral thermal slide technique[26] to remove the sapphire carrier substrate, as shown in Fig. 2B. Any residual PMMA and PPC are cleaned in acetone. BLAST is now complete and standard lithographic processing can now continue.

Figure 3 shows two examples of wafer-scale BLAST transfer of GaAs (Fig. 3A) and GaN (Fig. 3D) μLEDs from their native substrates to target Si substrates. The transfer process is performed across the entire 4"(GaAs) or 2"(GaN) wafer with high yield. The current and light emission as a function of voltage of the μLEDs before and after transfer are shown in Fig 3B(E) for GaAs (GaN), and the spectral characteristics are shown in Fig. 3C(F). These characteristics are nearly identical before and after transfer, demonstrating that BLAST does not damage the devices.

Figure 4A demonstrates the alignment accuracy of BLAST. Complementary alignment marks consisting of crosses and vernier structures were pre-fabricated on the native and target substrates before transfer; every alignment mark on the μLED substrate has a corresponding acceptor alignment mark fabricated on the target Si wafer (Fig. 4A, left). These marks both guide alignment and are used to quantify alignment accuracy (see Materials and Methods). Figure 4A (right) shows one such set of alignment marks; examples at multiple locations are shown in Fig. S6. The alignment accuracy is found to be approximately 1μm across the entire chip.

Figure 4B illustrates the high yield of BLAST for a variety of μLED geometries. We quantify the transfer yield by counting the missing μLEDs in a randomly selected 1 mm by 1 mm region. For both GaAs and GaN we obtain yields > 99.9% (Fig. 4B left and Fig. S5). If needed, the SU8 adhesion layer can also be left out of the process. For μLEDs transferred onto a bare silicon substrate, the yield is approximately 95% (Fig. 4B, right). In Figs. 4C, D, we show high yields for transferring μLEDs of varying sizes and heights. In Fig. 4C, μLEDs with lateral dimensions ranging from 10μm to 100μm are transferred with high alignment accuracy and nearly 100% yield. Figure 4D and Figure S7 show the transfer of μLEDs with heights ranging from 2–9μm. Overall, we show that both thin and relatively thick μLEDs of a variety of lateral dimensions are transferred with high alignment accuracy and yield.

To demonstrate the full capabilities of BLAST and its ability to integrate into an overall process flow, we fabricate a broadband optical upconverter (Figs. 1 and 5) that requires both materials integration and additional processing after transfer. Optoelectronic devices that convert low-frequency incident photons into higher frequency luminescent emission have gained broad interest in fields ranging from bio-sensing and infrared imaging[*27-31*]. Previous efforts mainly employed two approaches, both relying on the absorption of two photons to create a single photon of higher energy. The first, upconversion nanoparticles[*28-30*], work by nonlinear anti-Stokes emission[*32*] of two photons through a midgap state associated with a dopant. The second combines two GaAs photodiode junctions and AlGaInP light emitting diode junction[*31*] in a single vertical GaAs heterostructure stack.

Our micro-upconverter, by contrast, is made by integrating Si photovoltaic cells (PVs) in series with a GaN μLED (Figs. 1E, F, and 5). This approach allows separate tuning of the absorber and emitter and an arbitrarily large amount of upconversion to be realized by simply adding more PVs in series. For the devices shown, combining 6-12 Si PVs in series, each PV generates approximately 0.6 V, and generating 3.6-7.2 V, depending on the device. This is sufficient to drive the ~ 4 V threshold voltage GaN μLED (Figs. 3B,C) to emit blue light. The process produces thousands of micro-upconverters per wafer (Fig. 5A) and can explore many different device layouts on a single fab run.

This upconverter works for a broad range of excitation energies, from near-infrared to the visible, as shown in Fig. 5B, with the minimum incident photon energy set by the bandgap of the Si (~ 1.1 eV, or 1100 nm). This is in contrast to upconversion nanoparticles, which have relatively narrow absorption spectrum. Furthermore, the output light can reach to the blue. In contrast, tandem upconversion devices have limited emission wavelength because of the narrowly tunable bandgap of AlGaInP material. Overall, the heterogeneous integration approach used here allows us to independently tune the absorption and emission properties without materials constraints.

The upconversion efficiency is determined by the product of PVs' conversion efficiency (~ 20%) and μLED's light emitting efficiency (1-2 %)[*33,34*]. The overall upconversion power efficiency is therefore 0.2-0.4% (Fig. 5C). This is comparable to upconversion nanoparticles (typically ~0.01%-1%)[*35*] and the GaAs tandem upconverters (~ 1.5%)[*31*] and can be further improved by increasing GaN LED's light extraction efficiency and Si PV's light absorption efficiency. Surface roughening[*36,37*], metal reflection layer[*38*], micro structure array[*39,40*], transparent contact layer[*41*], and photonic crystals[*42*] can increase the efficiency of the GaN LED by up to ten times and the Si PVs by a factor of 2-3, leading to projected overall efficiencies > 10%.

The BLAST transfer process can be used for a wide variety of complex systems and materials, as shown in Fig. 6 and Fig. S10. Shown in Fig. 6A is the heterogeneous integration of GaAs vertical-cavity surface-emitting lasers (VCSELs) with foundry CMOS circuits. The VCSELs are fabricated on a GaAs heterostructure substrate, and then precisely integrated with CMOS circuits fabricated by a commercial vendor (XFAB-180 nm process). The resulting optical wireless integrated circuits (OWiCs) are powered by light (through the Si PVs). Once activated, each OWiC has programmability (through the integrated circuits) and can communicate to an external reader by VCSEL emitting digital optical signals (Fig. 6C). OWiCs have broad applications in sensing and identification[*7,43,44*], and their mass-fabrication is made possible by BLAST.

A second example is the aligned transfer 2D van der Waals materials shown in Fig. 6B. Stacked 2D materials have attracted major interest in the past few years for their unique electronic, thermal, and optical properties[*45*]. Here we show large-scale mass production of stacked graphene with precise spatial control (Fig. S13A). Graphene is grown on a copper host substrate by CVD. It is patterned on copper and then transferred onto a $SiO_2$/Si target substrate. This process is then repeated to form the stacked bilayer graphene devices shown in Figs. 6B, D. This process should be compatible with any van der Waals material, and makes possible mass manufacture with precise spatial and rotational control. We also show in Fig. S10 the transfer of Si PV devices onto a target glass substrate, relevant for applications in see-through circuitry[*18*].

We have demonstrated BLAST, a wafer scale transfer technique that can integrate photonic and electronic devices/circuits from different substrates onto a single platform suitable for additional lithographic processing. We showed its use with a variety of native substrates (GaAs, sapphire, SOI, Cu), target substrates (Si, $SiO_2$/Si, fused silica, and CMOS circuits on SOI) and optical devices transferred (GaN μLEDs, GaAs μLEDs, Si PVs, VCSELs, and graphene). BLAST is simple and scalable and has the potential to be applied on a vast range of materials systems, helping to make heterogeneous integration a standard fabrication tool in optoelectronic systems.

**Materials and Methods**

**GaAs μLED fabrication:** The GaAs heterostructure (AlGaAs/GaAs multiple quantum wells, cladding layers, contact layers and AlGaAs sacrificial release layer) used for producing LEDs was epitaxially grown on a 4inch GaAs wafer by a commercial vendor (Matrix Opto. Co., Ltd). Citric acid:$H_2O_2$ (20:1) is used to wet etch the heterostructure to expose the n-type GaAs layer. 2nm Ti and 9nm Pt are deposited with DC sputtering to form the p-contacts. Ni/Ge/Au (3nm/6.4nm/50nm) are deposited via e-beam evaporation and then annealed at 425°C for 1min under 20 sccm Ar flow to form the n-contacts. μLEDs' lateral geometry is defined by etching down to the AlGaAs release layer with citric acid:$H_2O_2$ (10:1).

**GaAs μLED transfer:** Prior to transfer, 2μm PMMA (495 PMMA A11, MicroChem) is spin-coated on the μLEDs as the protection layer. 10μm PPC (30 wt% in anisole) is spin-coated on a sapphire carrier wafer. Bonding the μLEDs coated by PMMA with the sapphire carrier wafer coated by PPC is carried out by a hot press under the temperature and pressure of 135°C and 1.5bar. The GaAs wafer and sapphire wafer stack is placed into citric acid and hydrogen peroxide mixture (4:1) to each away the GaAs bulk substrate followed by dipping into 100:1 $H_2O$:HF to etch away the AlGaAs release layer. By using an alignment system, the μLEDs are aligned with the target substrate and they are brought into contact, heating up the stack to 100°C. Afterward, we bring the stack to a hot press, keeping pressing at 0.7 bar pressure and 155°C for 15-30min. We then melt the PPC on a hotplate at 160°C allowing the removal of the sapphire carrier substrate. In the end, we remove the PPC and PMMA residual in acetone.

**GaN μLED fabrication:** The GaN heterostructure used for producing LEDs was either epitaxially grown on a 2inch non-patterned sapphire wafer (for demonstrating 2inch wafer-scale GaN μLED transfer, by University Wafer) or a 4inch patterned sapphire wafer (for fabricating all the other GaN μLEDs in the paper, by Xiamen Powerway

Advanced Material Co., Ltd.). $Cl_2$ based inductively coupled plasma (ICP) etching is employed to expose the n-type GaN layer. The deposition of Pd/Au (50nm/50nm) n-contacts and Ti/Au (25nm/100nm) p-contacts is carried by e-beam evaporation. The contacts are annealed at 660°C for 1min under Ar environment. µLEDs' geometry is defined by etching down to the sapphire substrate through $Cl_2$ based ICP etch. Even though PMMA and PPC are not light-sensitive, intense excimer laser light can still burn the polymer layer. To prevent that, we deposit a layer of Cr/Au (50nm/150nm) surrounding the µLEDs as the laser shielding layer.

**GaN µLED transfer:** We spin ~5µm PMMA (950 PMMA A11, MicroChem) and attach the µLED to a sapphire wafer coated with ~10µm PPC (30 wt% in anisole) using a hot press under the temperature and pressure at 135°C and 1.5bar. Laser lift-off is employed to delaminate the sapphire substrate (193nm ArF excimer laser, beam size: 220µm*220µm; beam energy density: 2000mJ/cm$^2$). The stack is then placed on a 45°C hotplate to melt the gallium and the sapphire native substrate is thus removed. Following that, the Ga residual is etched away in $HCl:H_2O(5:1)$. By using an alignment system, the µLEDs are aligned with the target substrate and they are brought into contact, heating up the stack to 100°C. Afterward, we bring the stack to a hot press, keeping pressing at 0.7 bar pressure and 155°C for 15-30min. We then melt the PPC on a hotplate at 160°C allowing the removal of the sapphire carrier substrate. In the end, we remove the PPC and PMMA residual in acetone.

**Si PVs fabrication and transfer:** We utilize spin-on glass to dope the SOI substrate's device layer to form PN junction layer. N-contact layer is exposed by HBr based ICP dry etch. PV mesa is outlined by HBr based ICP dry etch. The contacts and interconnects are deposited via DC sputtering of Ti/Pt. The Si PV chip is then coated with 2µm PMMA and bonded to a sapphire carrier wafer. Instead of wet etch and laser lift-off we applied to the µLED substrate removal, we employ Bosch deep reactive-ion etching to remove the bulk silicon substrate, and the etch stops at the $SiO_2$ BOX layer. The $SiO_2$ layer is etched in BOE (6:1). Following that, fused silica substrate is prepared as the target substrate with 500nm SU8 coating. The Si PVs are then attached to the fused silica substrate by hoptree at 155°C. The carrier wafer, PPC and PMMA are subsequently removed as the µLED transfer process.

**Graphene patterning and transfer:** Monolayer graphene is grown on copper via a commercial vendor (Grolltex). Ti/Pt alignment marks are deposited on graphene/copper via sputtering. Following that, we pattern the graphene via RIE $O_2$ plasma etch. The bilayer graphene is formed by aligned transfer with the assistance of the pre-deposited Ti/Pt alignment marks (Fig. S12).

**Alignment marks:** There are fine alignment marks to indicate the displacement. Bottom bar array represents the alignment in the x-direction. Left side bar array represents the alignment in the y-direction. When x-direction is perfectly aligned, the donor's and acceptor's central bar pair should be aligned. If the first bar pair on the left (right) is alignment, that means there is +0.5μm (-0.5μm) shift. If the second bar pair on the left (right) is alignment, that means there is +1μm (-1μm) shift, and so on. Based on the same idea, we can tell the alignment in the y direction.

**Acknowledgments**: We thank M. Reynolds, S. Norris, S. Bharadwaj, and the Cornell Nanoscale Facility staff members for fruitful discussions.

**Funding:** This work was supported by the Cornell Center for Materials Research (DMR1719875), by the Air Force Office of Scientific Research (MURI: FA9550-16-1-0031), by the New Frontier Grants from the Cornell College of Arts & Sciences, by the Kavli Institute at Cornell for N anoscale Science, and by the Cornell IGNITE Innovation Acceleration program. This work was performed in part at the Cornell NanoScale Facility, a member of the National Nanotechnology Coordinated Infrastructure (NNCI), which is supported by the National Science Foundation (Grant NNCI-2025233).

**Author contributions**: Y.J., A.J.C., and P.L.M. conceived the project. Y.J. and A.J.C. designed and developed the fabrication procedure for the electrical and optoelectronic devices. Y.J performed the device fabrication and integration. Y.J and C.L.S. carried out the optical characterization. A.J.C and A.C.M conducted the CMOS circuit design and integration with VCSEL. Y.J. and P.L.M. wrote the manuscript with contributions from all authors.

**Competing interests:** A.J.C., A.C.M., and P.L.M. are the co-founders of OWiC Technologies, Inc., a company developing microscopic optical smart ID tags that are fabricated in part using the BLAST process. A.J.C., A.C.M., and P.L.M. are inventors on a patent application submitted by Cornell University that covers Optical Wireless Integrated Circuits (US Patent App. 16/947,626, 2021) that are manufactured in part using the BLAST process. All other authors declare that they have no competing interests.

**Data and materials availability:** All data needed to evaluate the conclusions in the paper are present in the paper and/or the Supplementary Materials. Additional data related to this paper should be addressed to Paul L. McEuen.

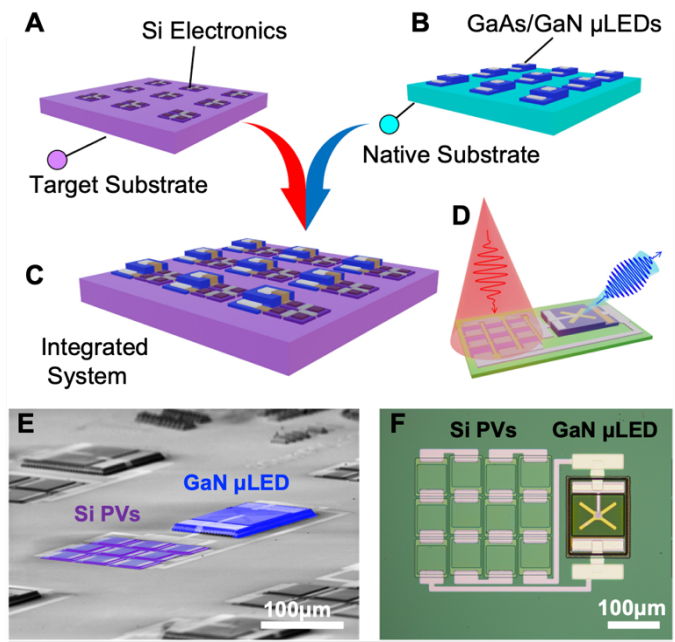

**Fig. 1. Heterogeneous Integration for Optoelectronics**

**(A)** Si electronics and **(B)** GaAs or GaN μLEDs are **(C)** integrated onto a single substrate to create devices such as **(D)** an optical upconverter. **(E)** False colored SEM image of an upconverter on Si. **(F)** Optical image of an upconverter on Si.

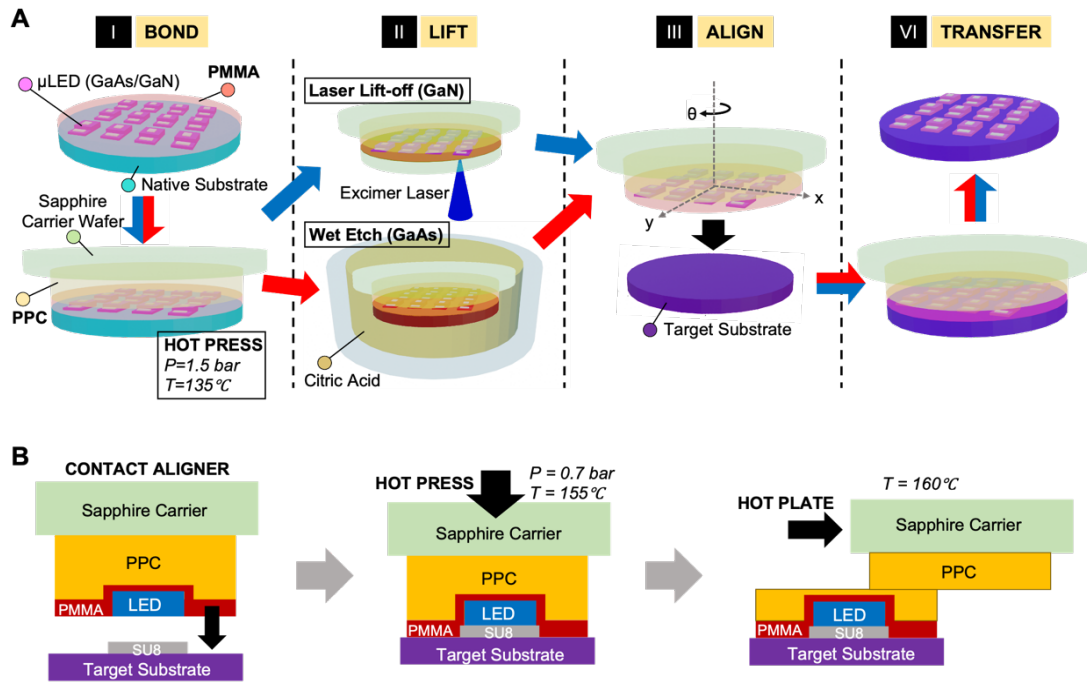

**Fig. 2. BLAST Process.**

(**A**) Schematic illustration of BLAST process. GaN and GaAs μLEDs are fabricated on their native substrates (sapphire and GaAs). To better demonstrate the consistency of the transfer process, we use a combined schematic to present both GaN and GaAs μLEDs. With prefabricated μLEDs, BLAST includes (*i*) *bonding* the μLEDs to a sapphire carrier wafer; (*ii*) *Lifting* the μLEDs from their native substrates (laser lift-off for GaN μLEDs and chemical wet etch for GaAs μLEDs); (*iii*) *Aligning* the μLEDs to the features on the target substrate; (*vi*) *Transferring* the μLEDs to the target substrate, with following carrier wafer & polymer removal. (**B**) Schematics of three key steps in the transfer process. They are alignment & temporary bonding to the target substrate, hot press bonding, and carrier wafer removal using a thermal slide technique.

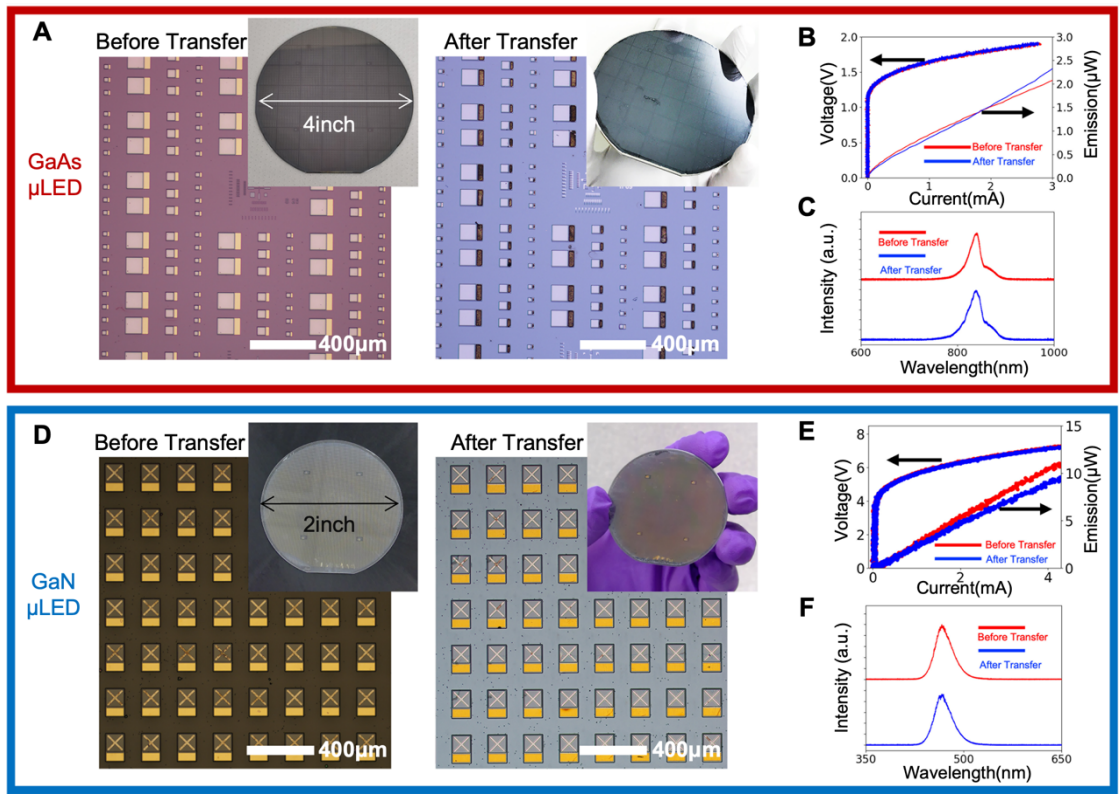

**Fig. 3. μLED Wafer-scale Transfer.**
(**A**) GaAs μLEDs before and after transferred onto a silicon wafer. (**B**) Current-voltage-emission curve of the GaAs μLED before and after transfer. (**C**) GaAs μLED emission spectrum before and after transfer. (**D**) GaN μLEDs before and after transferred onto a silicon wafer. (**E**) Current-voltage-emission curve of the GaN μLED before and after transfer. (**F**) GaN μLED emission spectrum before and after transfer.

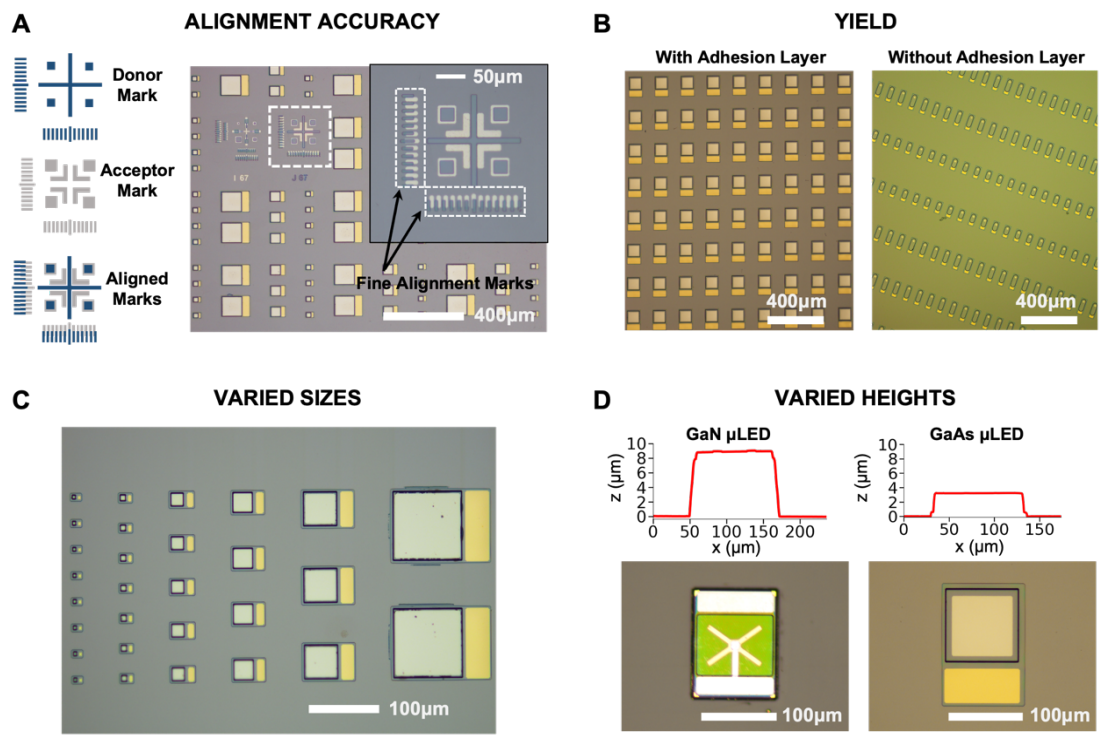

**Fig. 4. The Merits of BLAST Technique.**

**(A)** Schematic illustration of donor mark, acceptor mark and aligned marks (left). Optical image (right) of alignment marks showing our transfer has precision alignment (~1μm). **(B)** Optical image showing (left) the transfer has ~99.9% transfer yield with SU8 adhesion layer. Even without an adhesion layer like SU8, we still can get a transfer yield > 95% (right). **(C)** Optical image showing this transfer technique is able to transfer μLEDs in varied sizes from 10μm to 100μm. **(D)** Optical images and height profiles showing this transfer technique can transfer μLEDs in varied heights up to ~10μm.

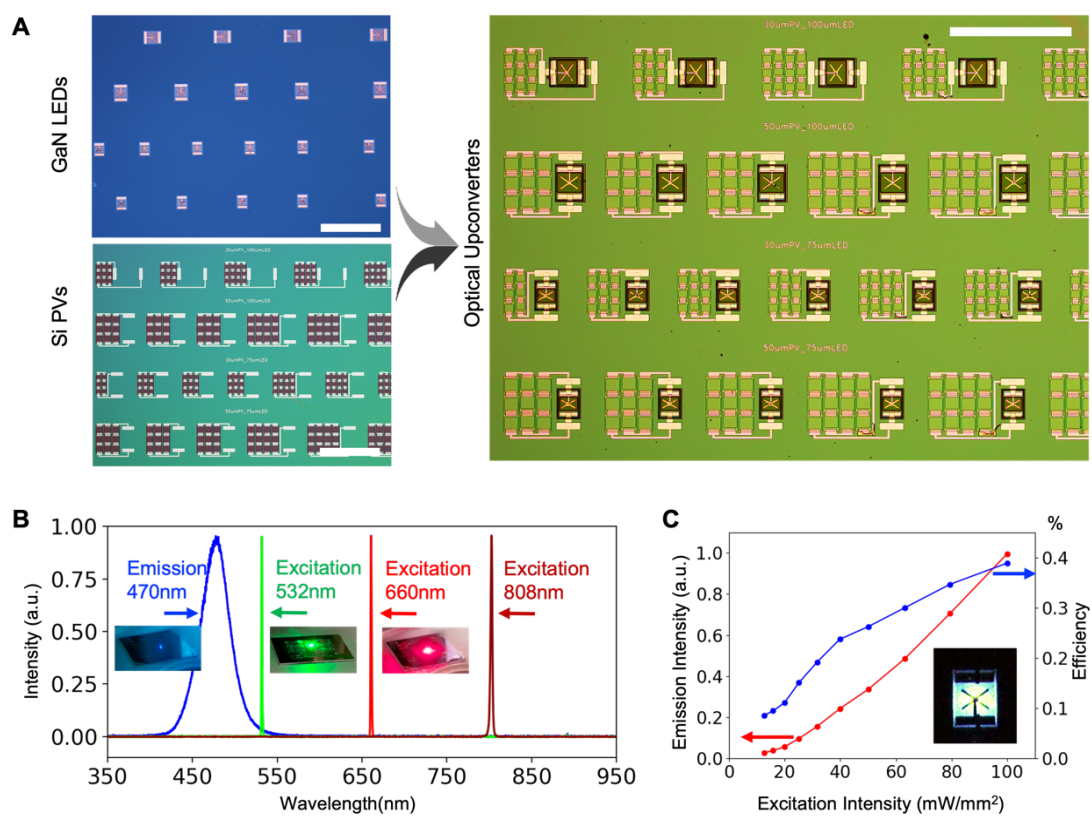

**Fig. 5. Optical Upconverter.**

**(A)** Optical images of upconverters fabricated in parallel (all scale bars are 500μm). The GaN LEDs are fabricated on sapphire and the Si PVs are fabricated on SOI. The broadband optical upconverter is made by transferring GaN LEDs from sapphire to SOI with subsequent lithographic processing. **(B)** Spectra of the excitation source (532nm laser, 660nm laser, and 808nm laser) and the upconverter's emission with a peak at 470 nm. **(C)** GaN μLED's emission power and conversion efficiency as a function of the illuminating laser's intensity. Inset is an optical image (with a 480nm optical bandpass filter) showing the upconverter emitting light.

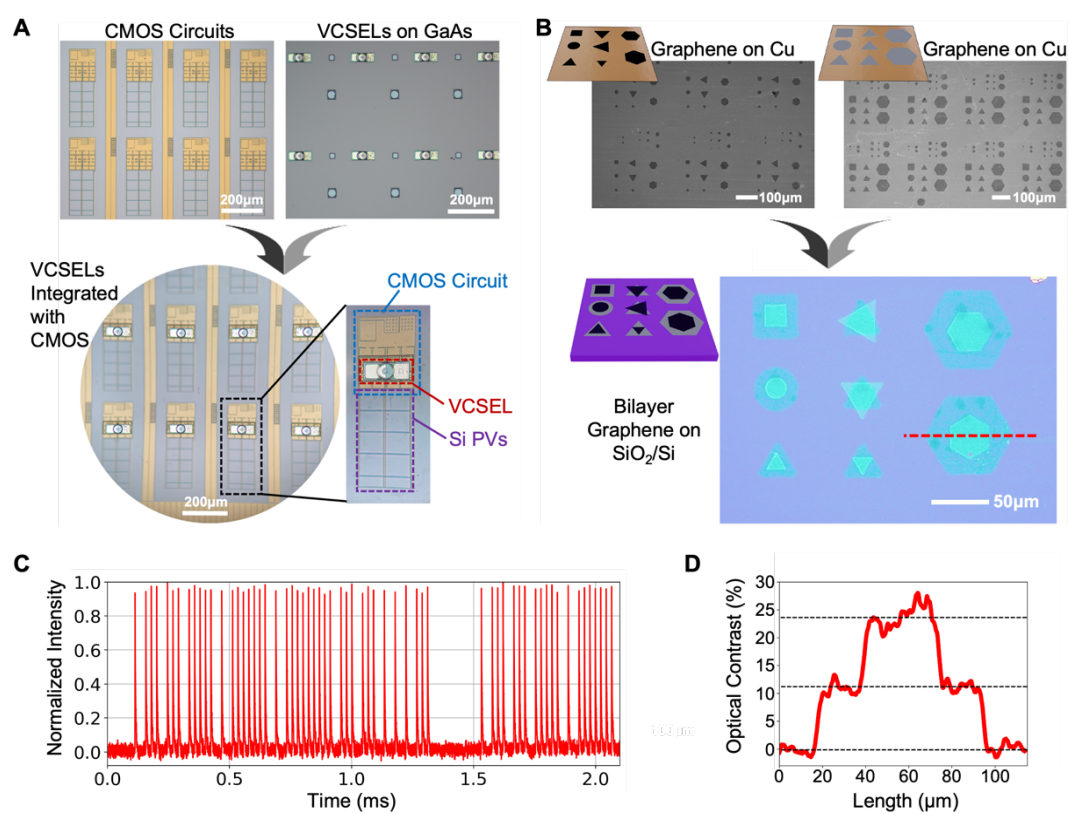

**Fig. 6. General Transfer.**

**(A)** VCSELs fabricated on a GaAs substrate are transferred and integrated with CMOS circuits. **(B)** Graphene is grown and patterned on Cu (SEM images, upper) and aligned transferred onto $SiO_2$/Si forming bilayer graphene stacks (optical image, lower). **(C)** Optical pulse trains (recorded by a photodetector) generated by the VCSEL for communicating to an external reader. **(D)** Optical contrast profile in the red channel of the CCD image along the line cut marked in **B**.

# BLAST: A Wafer-scale Transfer Process for Heterogeneous Integration of Optics and Electronics

Yanxin Ji[1], Alejandro J. Cortese[2], Conrad L. Smart[3], Alyosha C. Molnar[1,2,4], and Paul L. McEuen[2,3,4,*]

[1]Department of Electrical and Computer Engineering, Cornell University, Ithaca, NY 14853, USA

[2]OWiC Technologies, Ithaca, NY14853, USA

[3]Laboratory of Atomic and Solid-State Physics, Cornell University, Ithaca, NY 14853, USA.

[4]Kavli Institute at Cornell for Nanoscale Science, Cornell University, Ithaca, NY 14853, USA.

*Corresponding Author: Paul L. McEuen.     plm23@cornell.edu

**Table of Contents:**

**Upconverter Fabrication**

*GaN µLEDs Fabrication:* The GaN heterostructure is epitaxially grown on a patterned sapphire wafer (Xiamen Powerway Advanced Material Co., Ltd.). ICP RIE etching with $BCl_3/Cl_2/Ar$ is employed to expose the n-type GaN layer. The deposition of Pd/Au (50nm/50nm) n-contacts and Ti/Au (25nm/100nm) p-contacts is carried by e-beam evaporation. Contacts are annealed at 660°C in Ar for 1min. $SiO_2$ encapsulation layer is deposited via PECVD and patterned via $CHF_3/O_2$ based RIE etching. Ti/Pt (5nm/100nm) contacts protection layer is sputtering deposited and annealed at 350°C. µLEDs' geometries are defined by ICP RIE with $BCl_3/Cl_2/Ar$ to etch to the sapphire substrate. Following that, we deposit a layer of Cr/Au (50nm/150nm) surrounding the µLEDs as the laser shielding layer.

*Silicon PV Array Fabrication:* Silicon PVs are fabricated on SOI substrates with a 2µm device layer and a 500nm BOX layer. We spin-coat spin-on glass on the SOI substrate and anneal it at 1000°C to dope the device layer to form PN junction. The glass is etched away in 6:1 BOE. The n-contact layer is exposed by ICP dry etch (HBr). PV mesa is outlined by ICP dry etch (HBr). The contacts and interconnects are deposited via DC sputtering of Ti/Pt. We spin-coat 500nm SU8 and pattern it as the adhesion layer for µLED transfer.

*µLED and Silicon PV Array Integration:* We spin ~5µm PMMA onto the GaN µLEDs and attach the µLEDs to a sapphire wafer coated with ~10µm PPC using a hot press under the temperature and pressure at 135°C and 1.5bar. Laser lift-off is employed to delaminate the sapphire substrate (193nm ArF excimer laser, beam size: 220µm*220µm; beam energy density: 2000mJ/cm$^2$). The stack is then placed on a 45°C hotplate to melt the gallium and the sapphire native substrate is removed via mild mechanical force. Following that, the Ga residual is etched away in HCl: $H_2O$(5:1). By using an alignment system, the µLEDs are aligned with the PVs via alignment marks.

We drop water to the SOI substrate (to see clearly through PPC/PMMA since the patterns on the PSS substrate are copied to the PMMA layer) and bring the µLEDs and SOI substrate into contact. We then ramp the stage temperature from room temperature to 100°C for temporal bonding. Afterward, we bring the stack to a hot press, keeping pressing at 0.7 bar and 155°C for 15-30min. We then melt the PPC on a hotplate at 160°C allowing the slide removal of the sapphire carrier substrate. In the end, we remove the PPC and PMMA residual in acetone. After transfer, we deposit 300nm $SiO_2$ via PECVD as the dielectric pinning layer and pattern it via $CHF_3/O_2$ based RIE etching. The interconnects between PV array contacts and µLED contacts are made by sputtering deposition with Ti/Pt (10nm/100nm).

**Upconverter Optical Measurements**

*Optical Measurements Setup:* 532nm (CW laser, MGL-III-532-200mW, Ready Laser), 660nm (laser diode, Thorlabs), and 850nm laser (laser diode, Thorlabs) were used to excite the upconverter. For input vs. output and upconverter efficiency measurements, we adapted the 660nm red laser and an upconverter with nine 30µm*30µm PV and one 75µm LED. The measurements were performed with an upright microscope (BX51W1, Olympus) for laser illumination and upconverter emission measurements. The setup is shown in Figure S9. The laser output firstly passes through a beam expander (Thorlabs) and then is collimated into the microscope. A dichroic mirror is placed in the microscope to reflect the red laser light into the back of a 20X objective (Olympus) to illuminate the upconverter. The blue light emitted by the upconverter is collected by the 20X objective and passes through the dichroic mirror. The blue light is further filtered by a bandpass filter (HQ480/40X, Chroma) and reaches a silicon photodetector (DET36A, Thorlabs). The current generated by the Si photodetector is amplified by a current preamplifier (1211, Ithaco) and recorded with a digital oscilloscope (5242D, Picoscope).

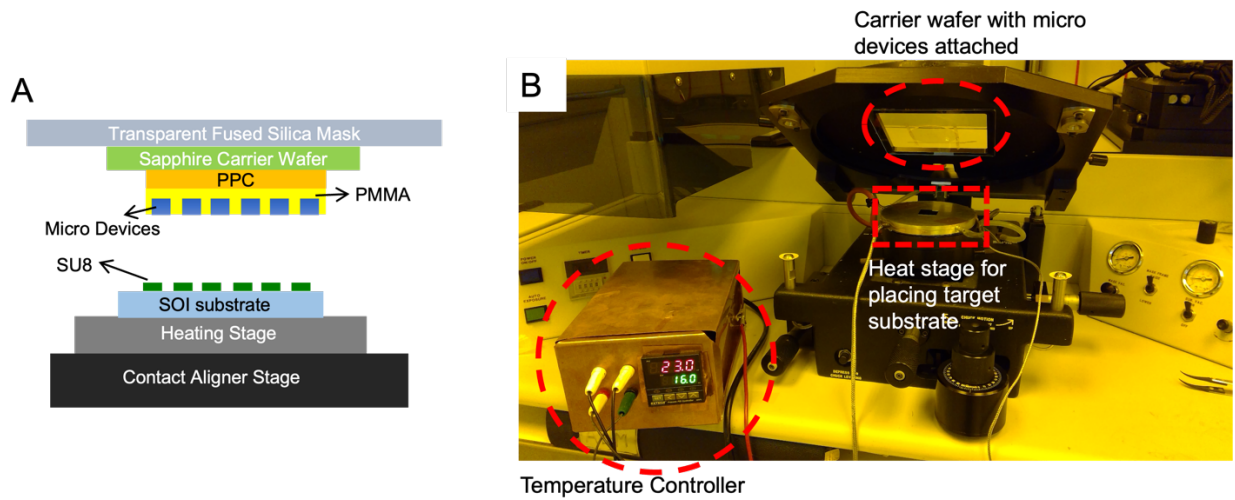

**Fig. S1.**

**(A)** Schematic illustration of the alignment system. **(B)** Photo showing key components of the alignment system.

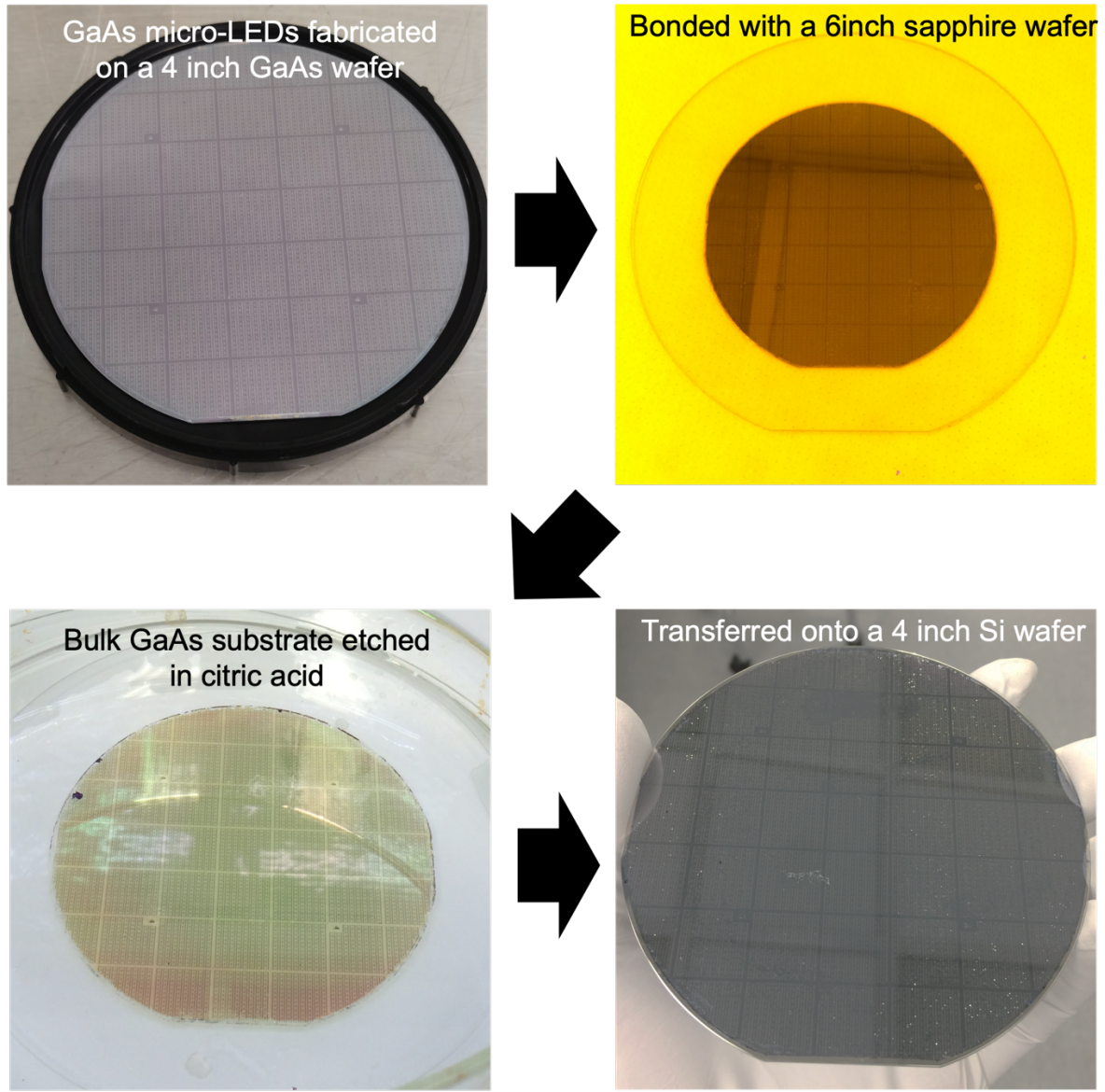

**Fig. S2.**

Photos showing the key steps of GaAs μLED wafer-scale transfer.

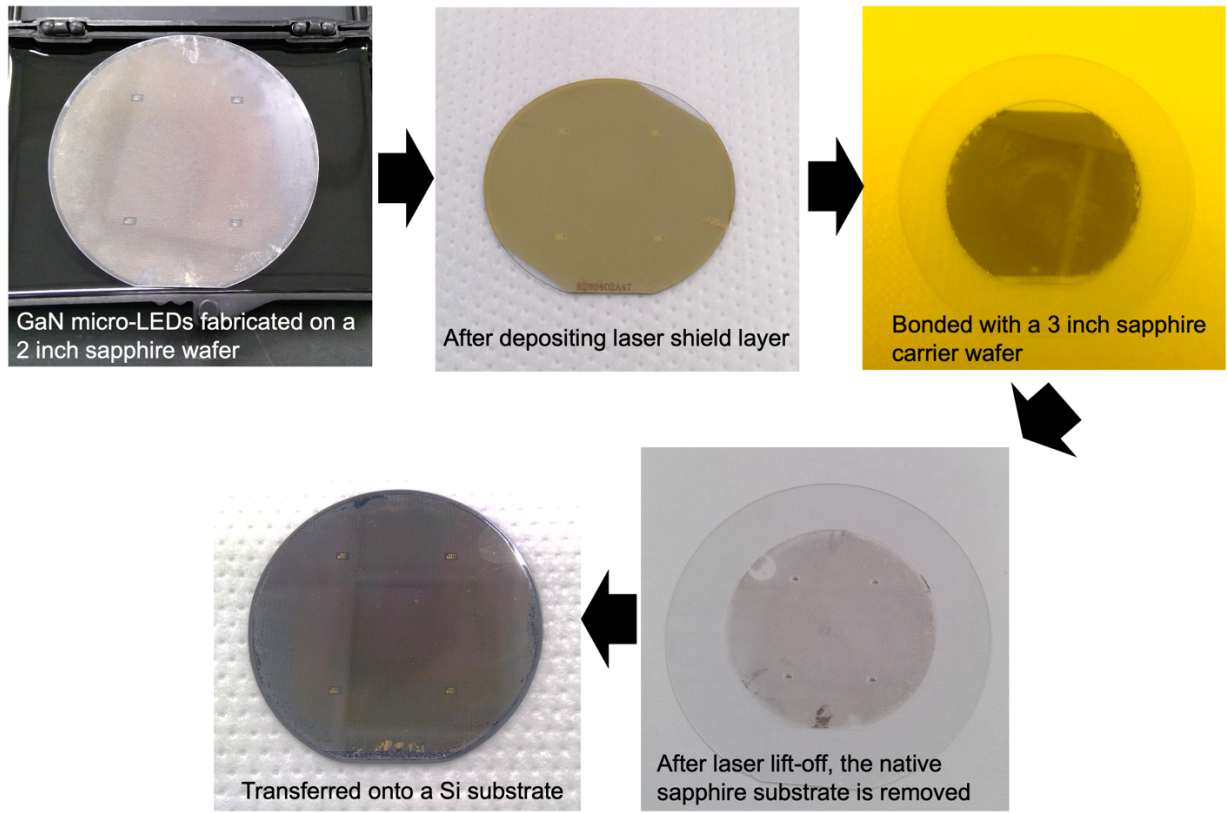

**Fig. S3.**

Photos showing the key steps of GaN μLED wafer-scale transfer.

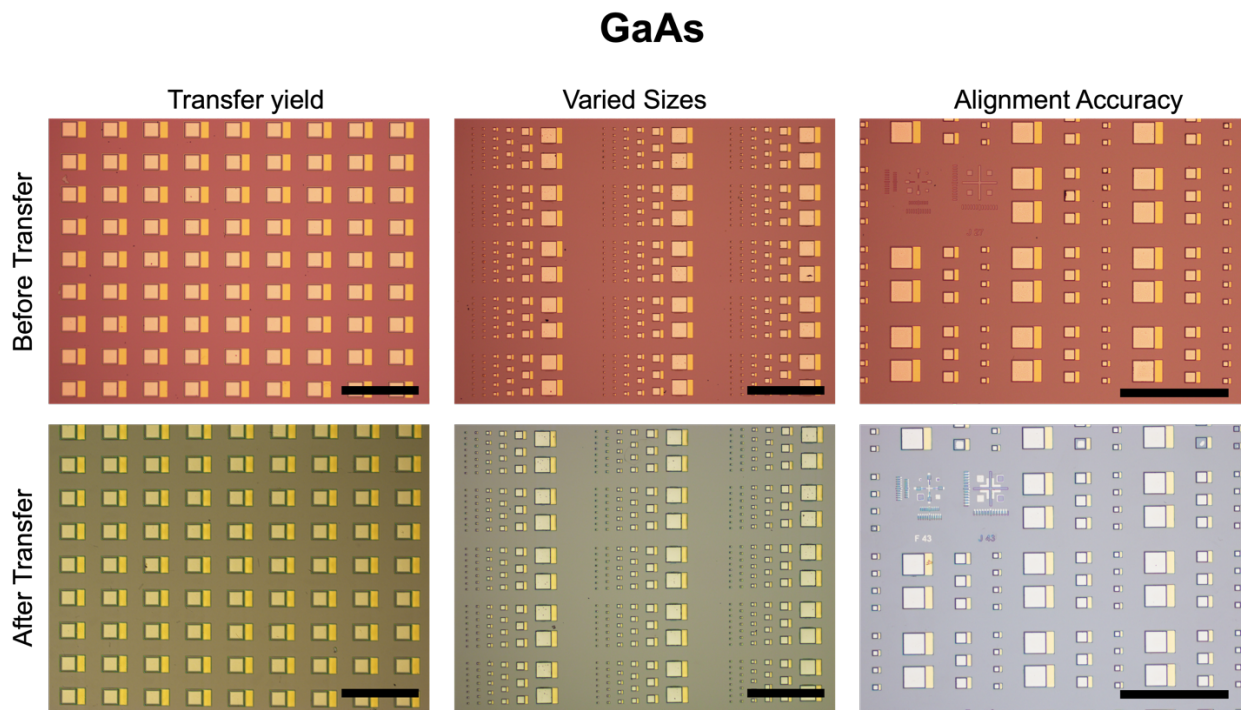

**Fig. S4.**

Optical photos comparing devices before and after transfer (GaAs). Scale bars are 500μm.

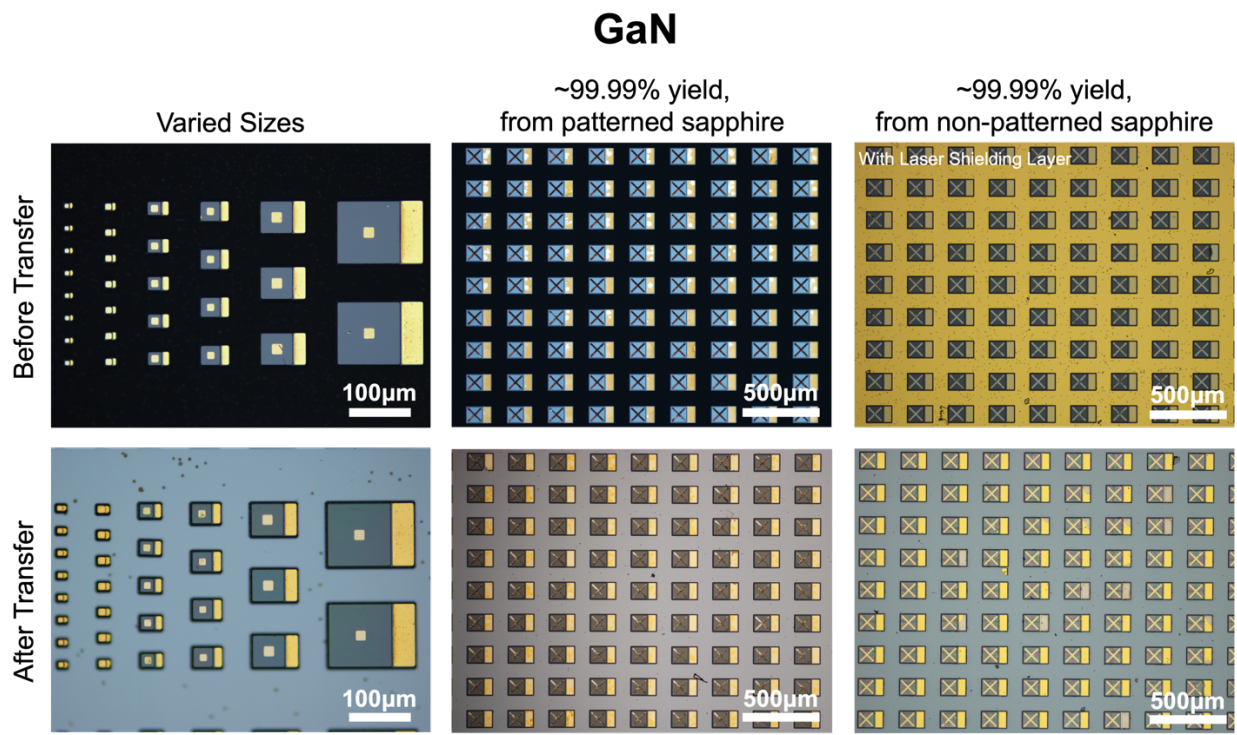

**Fig. S5.**

Optical photos comparing devices before and after transfer (GaN).

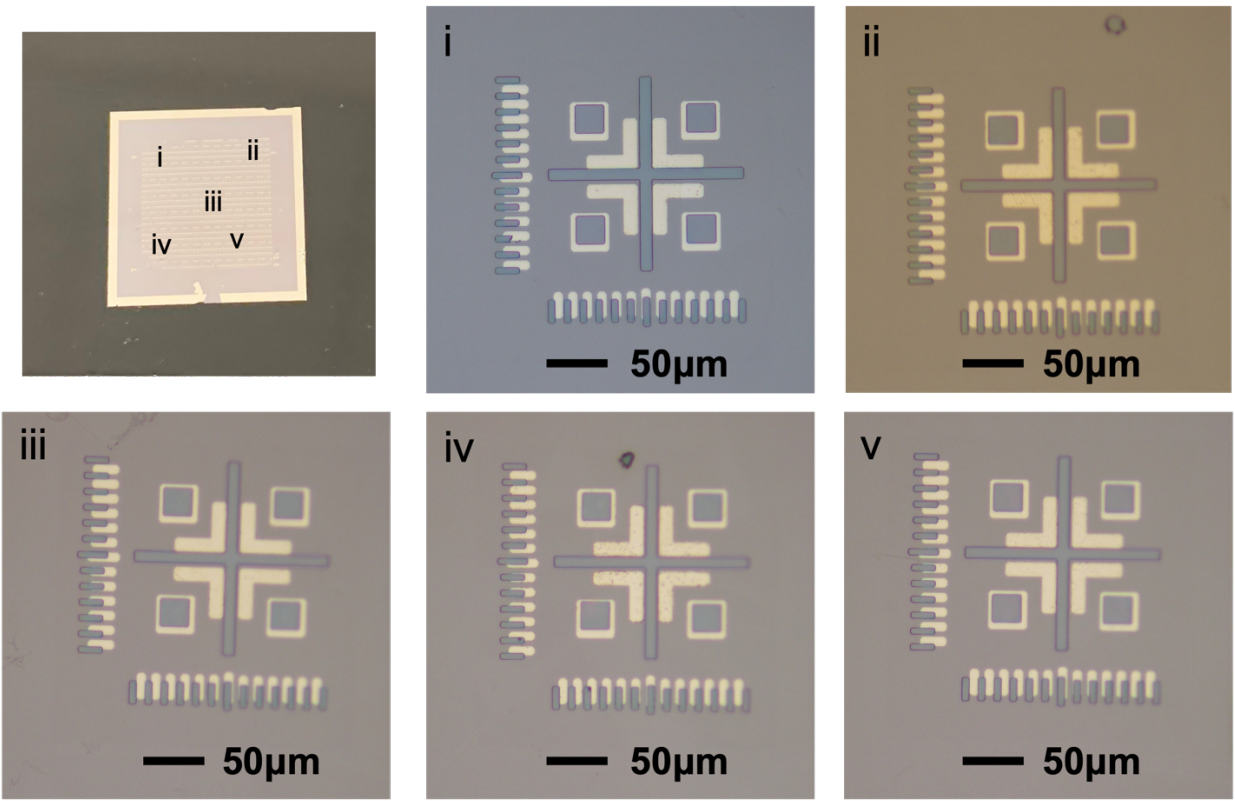

**Fig. S6.**

Alignment marks demonstrating ~1μm alignment accuracy across a 20mm*20mm chip.

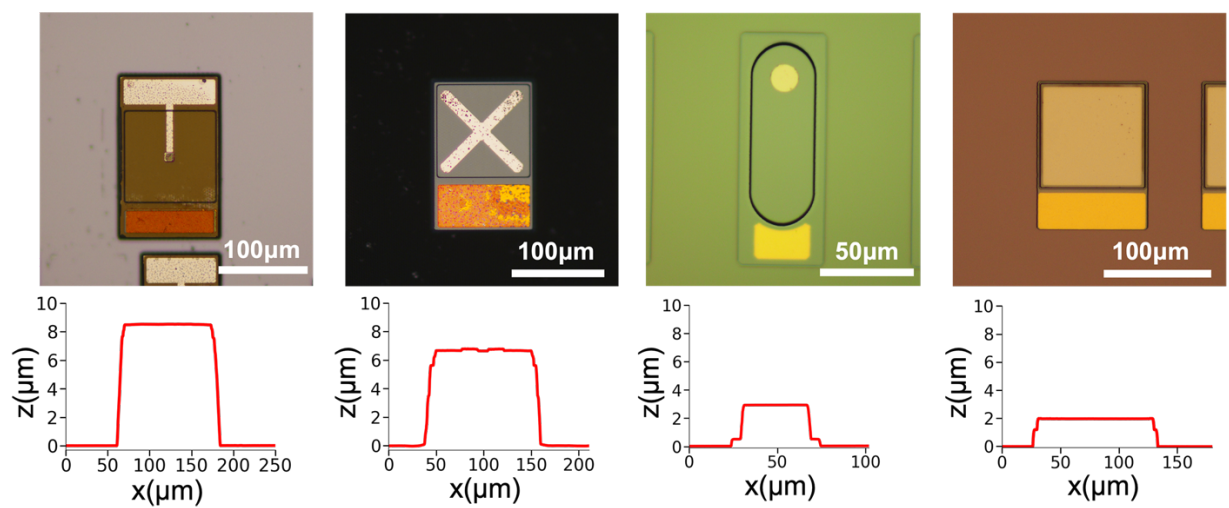

**Fig. S7.**

μLEDs with varied heights.

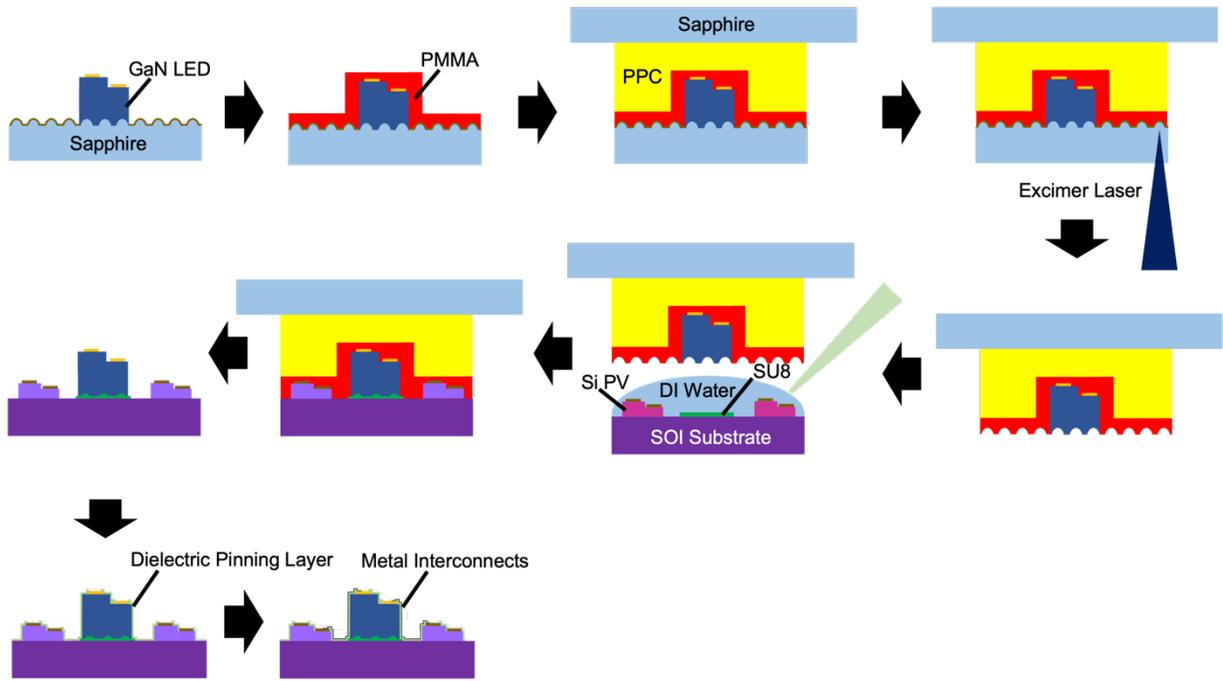

**Fig. S8.**

Schematic illustration of the upconverter integration process.

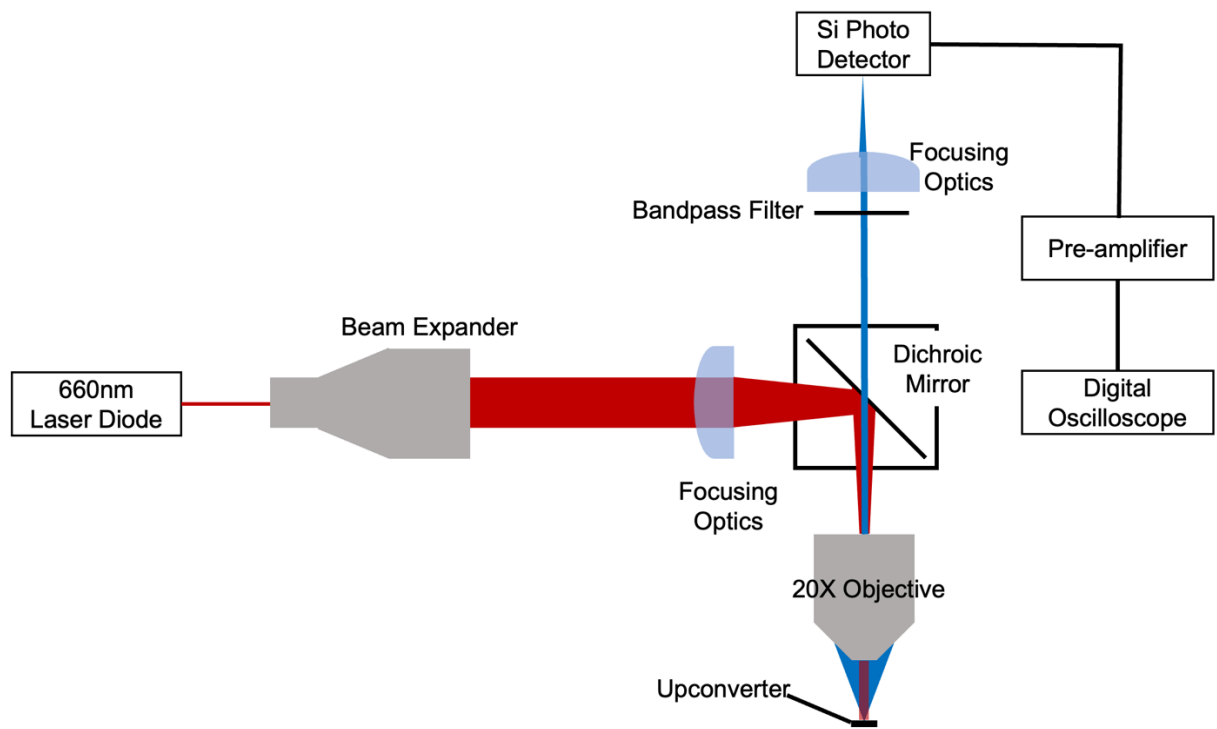

**Fig. S9.**

Schematic illustration of the optical setup for measuring upconverters.

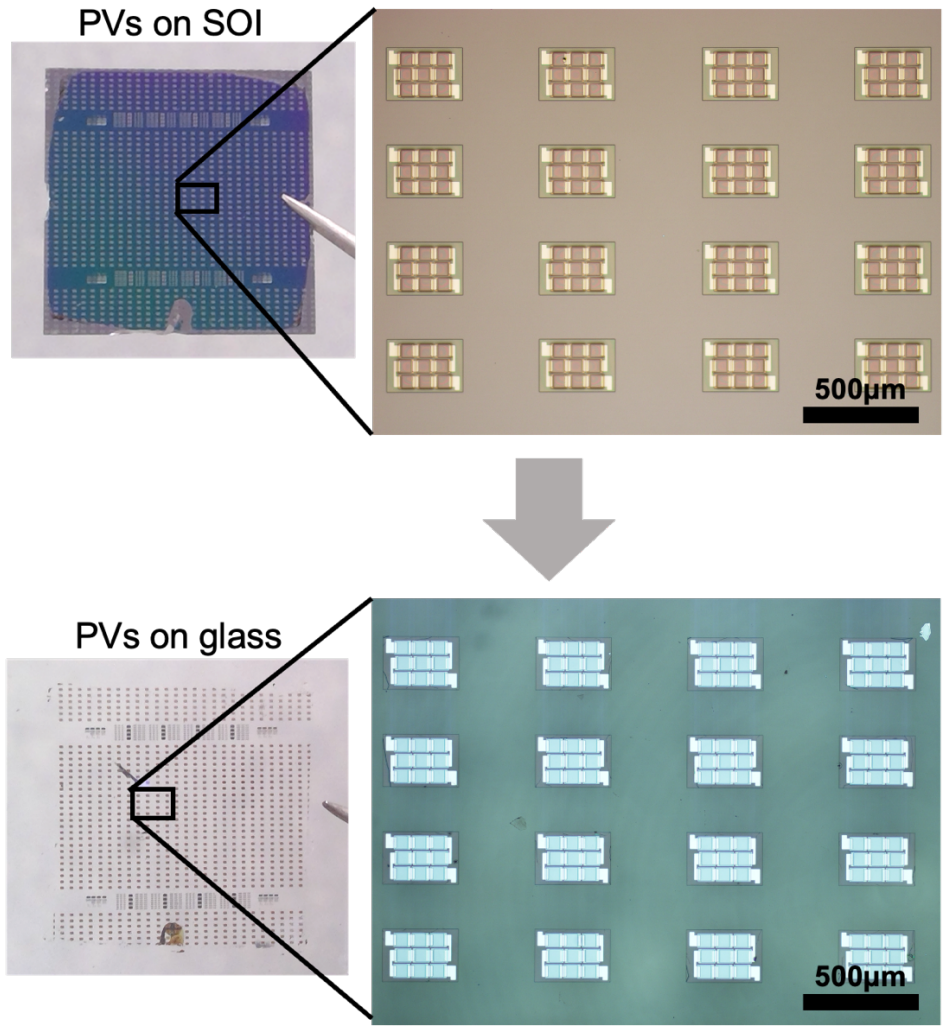

**Fig. S10.**

Silicon PVs are fabricated on an SOI substrate and then transferred onto a transparent glass substrate through BLAST.

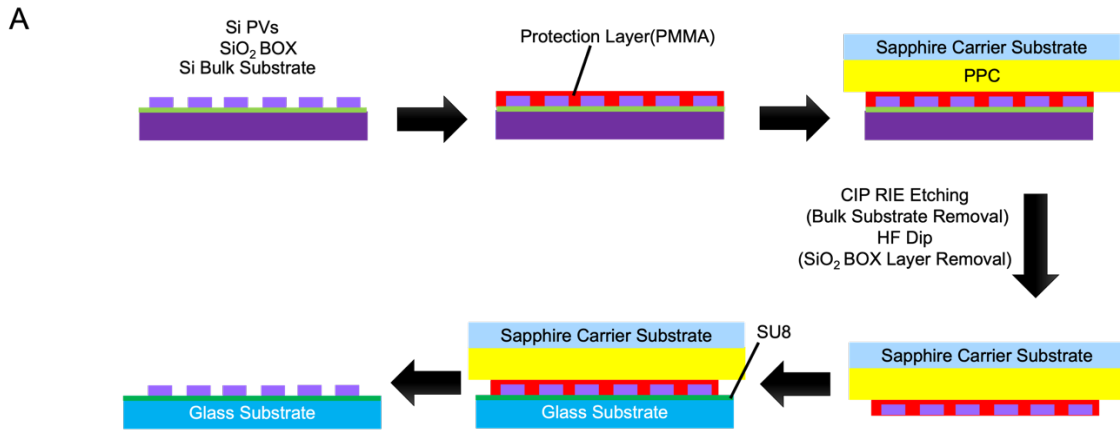

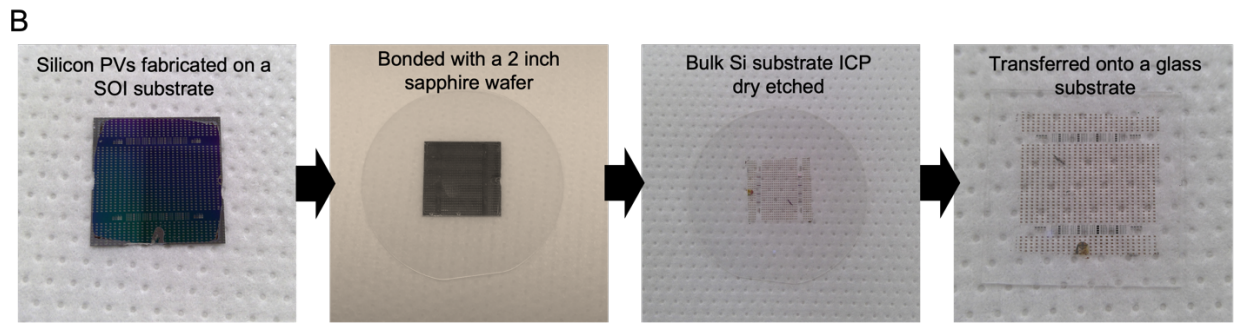

**Fig. S11.**

**(A)** Schematic illustration of the silicon PVs transferred from an SOI substrate to a glass substrate through BLAST. **(B)** Photos showing the key steps of the PV BLAST transfer process.

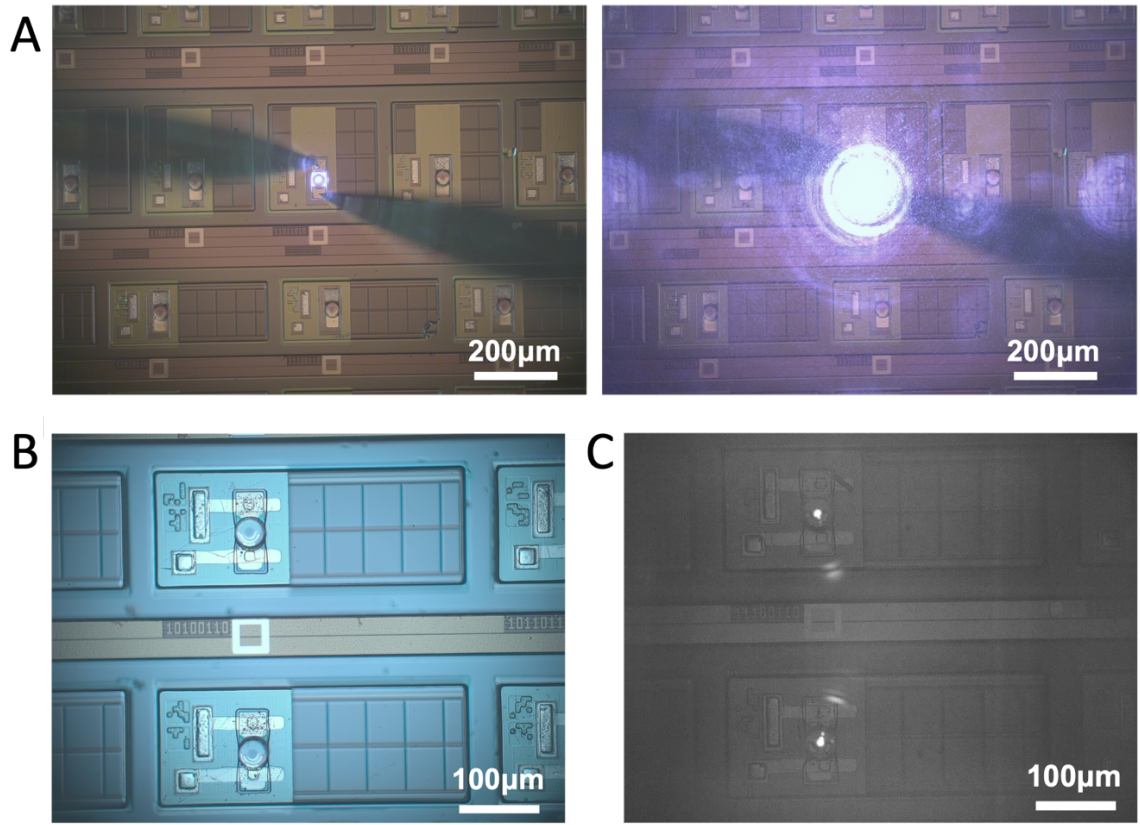

**Fig. S12.**

**(A)** With electrical probes on, VCSELs are tested to be functional after being transferred to the CMOS circuits. **(B)** Optical image showing VCSELs wired up with integrated circuits through post-transfer standard lithographic processing. **(C)** Optical image (with IR filter) of OWiCs in operation.

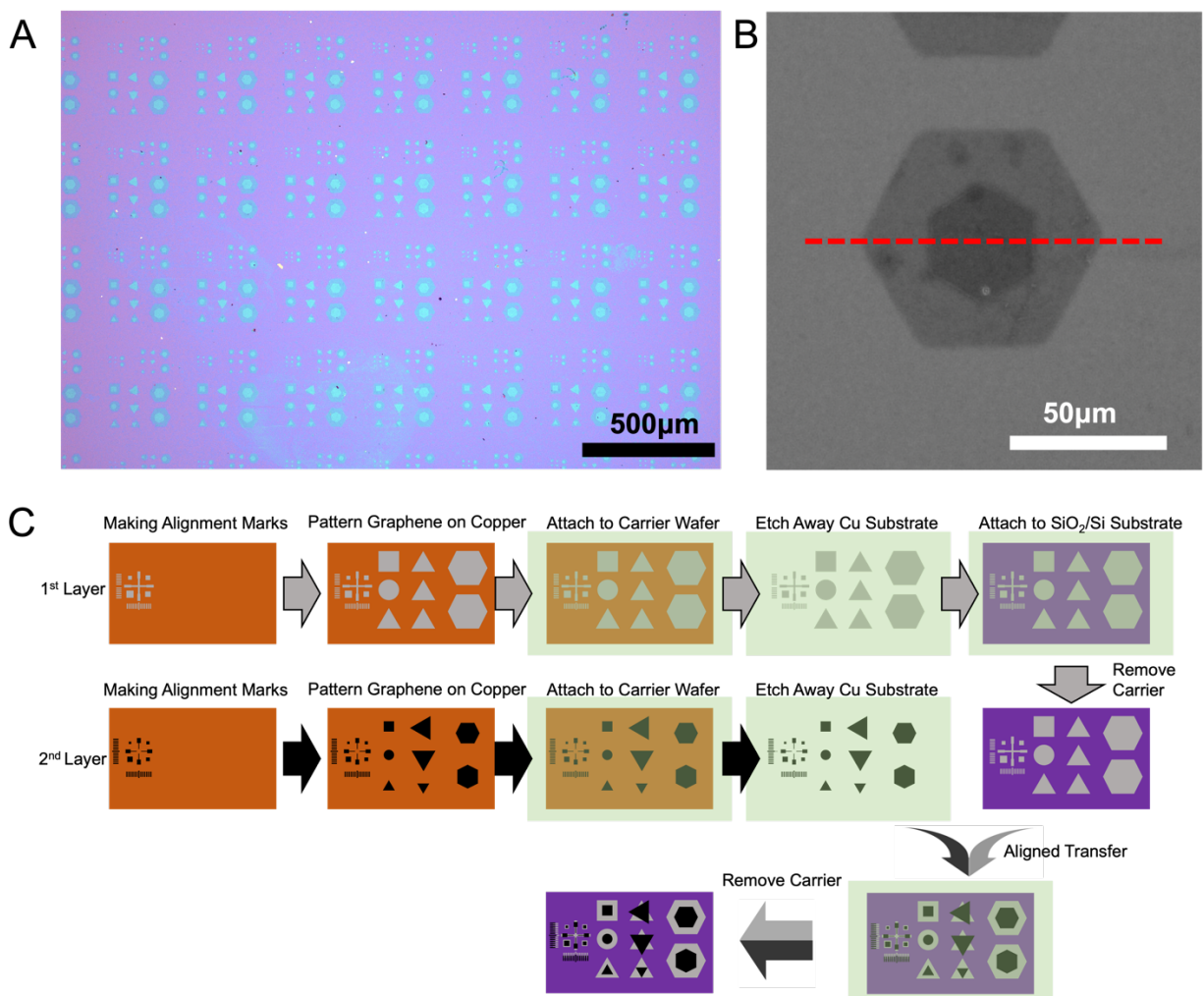

**Fig. S13.**

**(A)** Optical photo of large-scale bilayer graphene array. **(B)** The red channel of hexagonal bilayer graphene's CCD image. The red line cut corresponds to the optical contrast profile in **Fig. 6D**. **(C)** Schematic illustration of aligned bilayer graphene fabrication.